# Effect of Oxygen Adsorption on the Local Properties of Epitaxial Graphene on SiC (0001)

C. Mathieu[1], B. Lalmi[2], T. O. Mentes[3], E. Pallecchi[1], A. Locatelli[3], S. Latil[4], R. Belkhou[2] and A. Ouerghi[1]

1. CNRS- Laboratoire de Photonique et de Nanostructures, Route de Nozay, 91460 Marcoussis, France
2. Synchrotron SOLEIL, Saint-Aubin, BP48, F91192 Gif sur Yvette Cedex, France
3. Sincrotrone Trieste ELETTRA, Area Science Park, 34149 Basovizza, Trieste, Italy
4. CEA Saclay, DSM IRAMIS SPCSI, F-91191 Gif Sur Yvette, France

The effect of oxygen adsorption on the local structure and electronic properties of monolayer graphene grown on SiC(0001) has been studied by means of Low Energy Electron Microscopy (LEEM), microprobe Low Energy Electron Diffraction (µLEED) and microprobe Angle Resolved Photoemission (µARPES). We show that the buffer layer of epitaxial graphene on SiC(0001) is partially decoupled after oxidation. The monitoring of the oxidation process demonstrates that the oxygen saturates the Si dangling bonds, breaks some Si–C bonds at the interface and intercalates the graphene layer. Accurate control over the oxidation parameters enables us to tune the charge density modulation in the layer.

PACS number(s): 68.65.Pq, 61.48.Gh, 61.05.jh

## I. INTRODUCTION

Since its discovery, graphene has attracted tremendous interest and its unusual properties make it a promising candidate for future electronic and optic applications [1]. Along the view of designing graphene-based devices, many efforts are devoted nowadays to the achievement of large scale graphene patterning in a reproducible way with controlled structural quality [2]. Among the various ways to produce graphene [3], the growth of graphene layers on silicon carbide (SiC) is a very promising method for homogeneous large scale production with a high crystalline quality, as it has recently been demonstrated [4]. On the Si face, the first carbon layer presents a honeycomb structure but do not evidence the expected graphitic electronic properties [4, 5]. This is mainly due to the remaining one-third carbon atoms that are strongly bonded to the Si atoms of the substrate at the interface [6]. This first layer acts therefore as a buffer layer and promotes the growth of the following carbon layer, i.e. the first layer to behave, from an electronic point of view, as an isolated graphene sheet. However, the Si dangling bonds that remain trapped at the interface below the buffer layer are responsible for high intrinsic electron doping of graphene and hinder the carrier mobility. One way to overcome this limitation is to passivate the Si dangling bonds using dopants. As a matter of fact, it has already been demonstrated that intercalation of hydrogen [7], fluorine [8], gold [9] or lithium [10] can reduce this intrinsic doping and decouple the buffer layer transforming the interfacial layer into a purely $sp^2$ sheet of graphene. Recent experiments, performed using mainly LEEM microscopy, Core level photoemission (XPS), and Electron Energy Loss Spectroscopy (EELS), have demonstrated that the graphene π band of the buffer layer can be restored by the insertion of a thin oxide layer between the buffer layer and the SiC substrate [11]. However, none of these works is centered on the effect of oxidation on the electronic properties.

Here, we have followed, from local point of view, the structural and electronic modifications of epitaxial graphene, for different steps of the oxidation process. This has allowed us to finely study the transition of the buffer layer toward a partially decoupled graphene-like sheet, by a controlled *in-situ* exposition to oxygen. We demonstrate in the present letter that the oxygen can partially decouple the buffer layer from the substrate and hence reduce the intrinsic electron doping, which leads in turn to the partial transformation of the buffer layer into a graphene-like one. As a result, the oxygen reacts with the Si dangling bonds in the first place, reducing the charge transfer from the substrate to the graphene layer. Besides, our results indicate that molecular oxygen can intercalate between the carbon layers, with a (√3×√3)R30° pattern.

## II. EXPERIMENTAL DETAILS

The substrate used in these studies was semi-insulating on-axis SiC(0001). After polishing, the sample was exposed to hydrogen etching at 1600°C in order to remove polishing damages. The sample was further annealed at 1300–1400°C to grow the graphene layers under argon and silicon fluxes, as it favors the formation of large and homogeneous domains [12, 13]. During the graphitization process, the Ar partial pressure was kept below $P = 2 \times 10^{-5}$ Torr. The sample was then cooled down to the room temperature and transferred *ex-situ* from the growth chamber to the LEEM microscope chamber. Before measurements, the graphene sample was annealed at 600°C for 30 min in ultra-high vacuum, in order to remove the residual surface contaminations induced by the air transfer. The oxidation process has been carried out at 500°C under a pressure of ~$10^{-4}$ mbar of $O_2$, for a duration of 3 hours, in order to avoid oxygen etching of the graphene layer. Moreover, it has been shown that in this temperature range, the formation of Si–O bonds is favored [14]. This procedure has been repeated twice, after each the sample was fully characterized by means of LEEM, µARPES and µLEED.

The structural and electronic properties of the surface layers were probed by using the energy filtered LEEM-XPEEM microscope (SPELEEM III, Elmitec GmbH) operating at the Nanospectroscopy beamline of the Elettra storage ring in Trieste (Italy) [15]. This instrument enables to combine LEEM and XPEEM imaging with µLEED and µARPES, restricting measurements on an area of few square microns. Such analytical methods have been successfully used to characterize the local morphology, thickness, corrugation and electronic structure of single layer graphene films [16, 17].

## III. RESULTS AND DISCUSSION

Prior to the oxidation process, LEEM and µLEED measurements have been used to determine the as-grown graphene layer thickness. Several groups have used the electron (0,0) spot reflectivity oscillations to determine the number of graphene layers [16, 18, 19] in a LEEM setup. Using the same approach, we have identified predominately the growth of 1 and 2 monolayers (ML), with only a negligible amount of 3 MLs (Fig. 1(a)). As the buffer layer of the epitaxial graphene on SiC(0001) limits the carrier mobility [20], we focused our attention, during the oxidation process, only to areas corresponding to 1 ML (before oxidation) in the LEEM image, which are the most sensitive to the buffer layer. Figure 1(b) presents the electron reflectivity IV-curve recorded on the 1ML area before and after each step of the oxidation process; for a comparison a 2ML area reflectivity curve before oxidation is also presented, named hereafter pristine 2ML Gr/SiC. Before oxidation, the 1ML curve evidences only one minimum around 2.9 eV, typical of a single graphene layer [18]. After the first oxidation round, the minimum of the curve shifts of 0.4 eV toward the higher electron energy and a second minimum starts to develop around 2.2 eV electron energy (indicated by a black arrow on the Fig. 1(b)). After the second oxidation round no shift of the minimum is observed, but the second minimum becomes more pronounced and better defined around 2.0 eV. This latter curve starts clearly to resemble to a typical reflectivity curve of a pristine 2ML Gr/SiC, although the two minima do not appear at the same energies. It has been previously shown that when a graphene layer is exposed to hydrogen, the H intercalates between the substrate and the first carbon layer, which generates a complete transformation of a n-layer to a (n+1)-layer system [7, 21]. Therefore our findings support the transformation of the 1 ML into a 2 MLs subsequent of the oxygen intercalation at the interface during the oxidation process. For a pristine 2ML Gr/SiC, the reflectivity minima appear at higher electron energy, around respectively 1.7 and 4.4 eV and their relative intensity are almost similar (see bottom of Fig. 1(b)). However, for the second oxidation curve, the minima appear at 2.0 and 3.3 eV electron energy. This difference can be easily understood if we take into account that the buffer layer is only partially decoupled and that the residual Si–C bounds leads to the shift of the reflectivity minima, explaining these atypical reflectivity curves. However, we cannot exclude that the oxygen can locally intercalate between the two graphene layers; this can imply roughness, leading to the lost of coherency between the layers. This phenomenon might also explain the difference between the curves recorded after oxidation processes and the one obtained on a pristine 2ML Gr/SiC area.

To elucidate this phenomenon, µLEED patterns have been recorded averaging the signal over the whole sample's area within the field of view of the microscope. Figure 1(c) and 1(d) display diffraction patterns before and after two oxidation cycles, respectively. The pattern, observed in the Fig. 1(c), presents three contributions. The first one results from the unreconstructed hexagonal (1×1) structure, and is usually attributed to the signature of the SiC substrate. The second one, rotated by 30° with respect to the substrate contribution, arises from the sharp (1×1) graphene layer, and confirms the presence of a homogenous surface. The last one, which is characterized by isotropic (circular) 1/6 fractional spots, is attributed to the (6√3×6√3)R30° reconstruction. The latter superstructure, associated to the buffer layer, is clearly observed before oxidation, demonstrating the high structural order and good quality of the sample. After the second oxidation (Fig. 1(d)), the spots associated to the buffer layer are strongly suppressed, but are still detectable. This observation provides further evidence that the buffer layer is not totally decoupled, and that some Si–C bonds still are unsaturated. Combining the information obtained by the reflectivity curve and the µ-LEED after the oxidation, one may conclude at this stage that the oxidation process does not completely decouple the buffer layer from the substrate, as in the case of the hydrogenation.

The oxygen atoms penetrate the surface to saturate the Si dangling bonds and break some of the Si–C bonds, holding the buffer layer. For such intercalation/penetration through the graphene lattice, possible scenarios can be considered: either by the step edges, or through grain boundaries, this latter case being more probable, since it has already been observed in the case of hydrogenation [7]. However, the oxygen can also intercalate between the interface and 1 ML of graphene or between graphene sheets. Indeed, during our experiments, we found on one area, that the oxygen exposition produces new superlattice spots with the periodicity of (√3×√3)R30°, for which the rotation of 30° is relative to the graphene (1x1) cell, as indicated in the Fig. 2(a) by the green arrows. On this LEED image, this structure is superimposed to the (6√3×6√3)R30° superstructure which is strongly attenuated. This superlattice may induce a strong modification either in structural or in electronic properties of graphene and could be of great interest in order to tune the physical properties of carbon nanomaterials. Different hypothesis can be made to explain the appearance of the (√3×√3)R30° structure:

(i) Oxygen atoms can adsorb on step edges. In such a case, the C–C double bonds would be frozen, leading to a pattern of Clar sextets [22]. This reconstruction would lead to a (√3×√3)R30° structure. Therefore, if such a structure exists due to oxygen adsorption on edges, this should be also observable after exposing the sample to air, without any further surface treatment, which is not the case.

(ii) Starke *et al.* have observed a (√3×√3)R30°/SiC structure after exposing a SiC(0001) sample to the air, and have attributed this structure to the formation of $Si_2O_3$ species [23]. In this case the rotation of 30° is relative to the SiC (1x1) structure. In our case, the superstructure presents a rotation of 30° relative to the graphene one (R30°/Gr). The oxygen will react with the silicon dangling bonds, forming $Si_xO_y$ species. However, these species will never exhibit such a (√3×√3)R30°/Gr pattern.

(iii) The oxygen atoms can chemisorb onto the graphene layer, transforming the hybridization of the carbon atoms from $sp^2$ to $sp^3$. As it is well known, among the interesting properties of the pristine graphene, there is the reduced reactivity which makes it a highly inert material. This means that the oxygen can only adsorb on the graphene layer at defects. The LEED image of the clean and unexposed surface (Fig. 1(c)) proves the good crystalline quality of the graphene layer, implying that only few defects are present on the surface [12]. Moreover and as it will discussed hereafter, the ARPES measurements evidence a clear Dirac cone at the K point of the Brillouin zone after oxidation, which is also a fingerprint of the high graphene layer quality. If some oxygen had induced structural defects on the honeycomb lattice of the graphene layer, the Dirac cone would be strongly affected by the oxygen chemisorption.

None of these three scenarios can provide a satisfactory explanation of the observation of the (√3×√3)R30° pattern. To explain this superstructure, the alternative is to assume that oxygen is intercalated between the buffer layer and the upper graphene one, without chemical bonds with the graphene, as drawn in the sketch of figure 2(b). This scenario excludes the possibility to adsorb atomic oxygen since it will create necessarily a chemical bond, due to its high reactivity. Thus, the atomic oxygen atoms will never organize themselves on a periodic structure, with a $C_6O$ stoichiometry. Such covalently bonded O defect (e.g. epoxy) will locate randomly on the graphene layer and will strongly damp the electronic mobility, as it has recently been studied [24]. On the other hand, a scenario consistent with all the experimental observations is an intercalation of the molecular oxygen with a (√3×√3)R30° superlattice, which is common in intercalation compounds [25].

To progress further and to clearly evidence the effect of the oxidation on the sample, and thus its effect on the electronic structure, local ARPES measurements have been performed. Figure 3 displays the valence band structure around the K point of the first graphene Brillouin zone, before and after each step of the oxidation process. These 2D maps have been recorded over the whole sample's area within the field of view of the microscope, as it has been done in our μLEED experiment. On the clean unexposed surface (Fig. 3(a)), the linear dispersion of the π band can be observed [12, 26]. As it has been observed many times in the case of epitaxial graphene with the Si termination, the π bands form a cone, for which the π branches cross at the Dirac point ($E_D$) at –0.5 eV below the Fermi level ($E_F$) [12, 27, 28]. This energy shift $\Delta E$ ($\Delta E = E_D - E_F$) is known to be related to the charge transfer from the substrate to the graphene layers [29, 30], due to the Si dangling bonds. After exposure to $O_2$ (Fig 3(b)), the Dirac point shifts towards the Fermi level ($\Delta E = 0.3$ eV), proving that the electronic transfer from the substrate to the graphene layers is partially cancelled. This can be explained by the saturation of the Si dangling bonds with the oxygen atoms. This charge transfer is still detected after the second oxidation ($\Delta E = 0.2$ eV), but the shift toward the Fermi level is lightened in this case (Fig. 3(c)), suggesting a damping of the diffusion rate of oxygen atoms i.e. the main effect has been already reached during the first oxidation. At this stage (i.e. after second oxidation), the full width at half maximum (FWHM) of the Dirac cone branches increases from 0.10 before oxidation to ~0.14 Å$^{-1}$. The broadening of the Dirac cone branch can be attributed to the formation of a second π branch that cannot be solved within the energy resolution limit of the LEEM instrument (250meV). This new π branch might be associated to the appearance of a bilayer, due to the partial decoupling of the buffer layer. However, we cannot exclude that this might also be either an effect of local inhomogeneities or corrugation and defects in the graphene layer, induced by the oxidation process. This finding strongly supports the picture of partial decoupling of the buffer layer, as it has been already observed on a graphene bilayer [21, 27]. Nevertheless, these π branches are still clearly defined, proving that the oxidation does not affect the 2D structure of the graphene, corroborating in turn the oxygen intercalation process. The ARPES experiment confirms that the oxidation leads to saturate the Si dangling bonds with a partial decoupling of the buffer layer, along with an intercalation process of oxygen species. Using the linear dispersion of the density of states near the Dirac point, the charge carrier concentration (electron or hole) can be estimated, thanks to the classical relation: $|n|=(\Delta E)^2/[\pi(\hbar v_F)^2]$, where $\hbar$ is the Planck's constant, and $v_F$ the Fermi velocity, which was determined equal to $1.1\times10^6$ m.s$^{-1}$ in our measurements. The resulting electron density changes from n~$1.4\times10^{13}$ to $4.6\times10^{12}$ cm$^{-2}$ after the first oxidation, and decrease down to $1.9\times10^{12}$ cm$^{-2}$ after the second one. This decrease can be explained by two simultaneous phenomena. On the one hand, the Si atoms relying from the SiC substrate can be saturated. On the other hand, the oxygen present in the system can withdraw the exceeding charge of the material. The ARPES combined with the LEEM experiments demonstrate that the oxidation of the graphene layer reduces the intrinsic electron doping of this material. The LEEM oscillations and the shift of the Dirac point toward the Fermi energy allow controlling the charge density, which is one of the key to generate materials optimized for electronic device applications. Indeed, first transport measurements on 50×50 µm$^2$ Hall bar, fabricated from this oxidized sample, indicate a low electron doping n~$9\times10^{11}$ cm$^{-2}$ and show a signature of typical graphene monolayer, consistent with an interface layer not completely decoupled from the SiC substrate [31].

The combination of µLEED, LEEM and µARPES experiments gives a global picture of the structural and electronic graphene properties after the oxidation process. Our results converge towards the partial decoupling of the buffer layer. Figure 4 presents a scheme of the oxidation process as deduced from these results. As it is now well established on the SiC(0001) surface, the buffer layer presents some covalent bonds with the substrate, leading to Si dangling bonds, which affects the carrier density and the mobility of the graphene (Fig 4(a)). After the first oxidation, the oxygen diffuses and migrates under the buffer layer, react with the unbound Si atoms, and break some of the Si-C bonds holding the buffer layer to the substrate (Fig. 4(b)). However, all the Si-C bonds are not broken, since the (6√3×6√3)R30° can still be observed on the LEED pattern after the second oxidation. The (√3×√3) pattern, observed on a different area of the sample, suggests that the oxygen can intercalate between the interface and the first graphene layers, without creating covalent bounds with the graphene layers (Fig. 4(c)). One remaining question is how the oxygen penetrates through the graphene layer. At this annealing temperature (T=500°C), for which the migration might be facilitated, the oxygen atoms can either diffuse laterally through the step edges or vertically either through the (6√3×6√3)R30° buffer layer or through the few defects present on the graphene sheet, as it has already been theoretically proposed in the case of the Si out-diffusion, on epitaxial graphene [32].

## IV. CONCLUSION

In summary, we have demonstrated that oxidation process can be used to partially decouple the graphene layer grown on SiC form on its interface. The oxygen can diffuse through the graphene and buffer layer to saturate the Si dangling bonds and to break some Si–C bonds, and hence decouple partially the buffer layer. The intercalation of oxygen between these two layers has also been highlighted by the occasional presence of a (√3×√3)R30° pattern, observed by µLEED. These LEEM and µARPES experiments have opened up a route to modulate the charge density in graphene, which is of a prime importance for the electronic device applications.

**Ackowledgement:** This work was supported by the two French contracts ANR-2010-MIGRAQUEL and ANR-2011-SUPERTRAMP, and the RTRA Triangle de la Physique.

**Caption figures:**

**FIG. 1.** (Color online) (a) LEEM image of a clean and unexposed graphene/6H-SiC(0001) surface, with an electron beam energy of 2.50 eV. (b) Typical reflectivity curves extracted from the LEEM data before and after oxidation. A typical reflectivity curve of a graphene bilayer obtained before oxidation is also presented. The reflectivity curves are set from an electron beam energy of 1 to 10 eV. The local IV curves have been obtained by integrating the intensity signal on small selected are of LEEM image, while changing the primary electron energy. These curve correspond roughly to the classical (0,0) LEED spot IV-curve, but measured over small microsized area. (c)–(d) µLEED images, taken et 45 eV, before and after the second oxidation respectively. The images present the SiC, graphene and buffer structures, noted SiC, G and B respectively.

**FIG. 2.** (Color online) (a) µLEED pattern, obtained at 45 eV, of a ($\sqrt{3}\times\sqrt{3}$)R30° superstructure. (b) Schematic representation of the oxygen intercalation between graphene layers, which explains this superstructure detected in LEED.

**FIG. 3.** (Color online) (a)–(c) Band dispersion as a function of $k_{//}$ around the $K$ point of the first Brillouin zone, obtained by µARPES at hν=40 eV, performed before oxidation and after each oxidation step. The Fermi level and the Dirac point are superimposed on the images.

**FIG. 4.** (Color online) Sketch of the proposed oxidation process model. After the oxidation process, the oxygen penetrates the surface layer to saturate the Si dangling bonds and break some Si–C bonds and intercalates between the buffer and graphene layer.

**FIGURE 1**

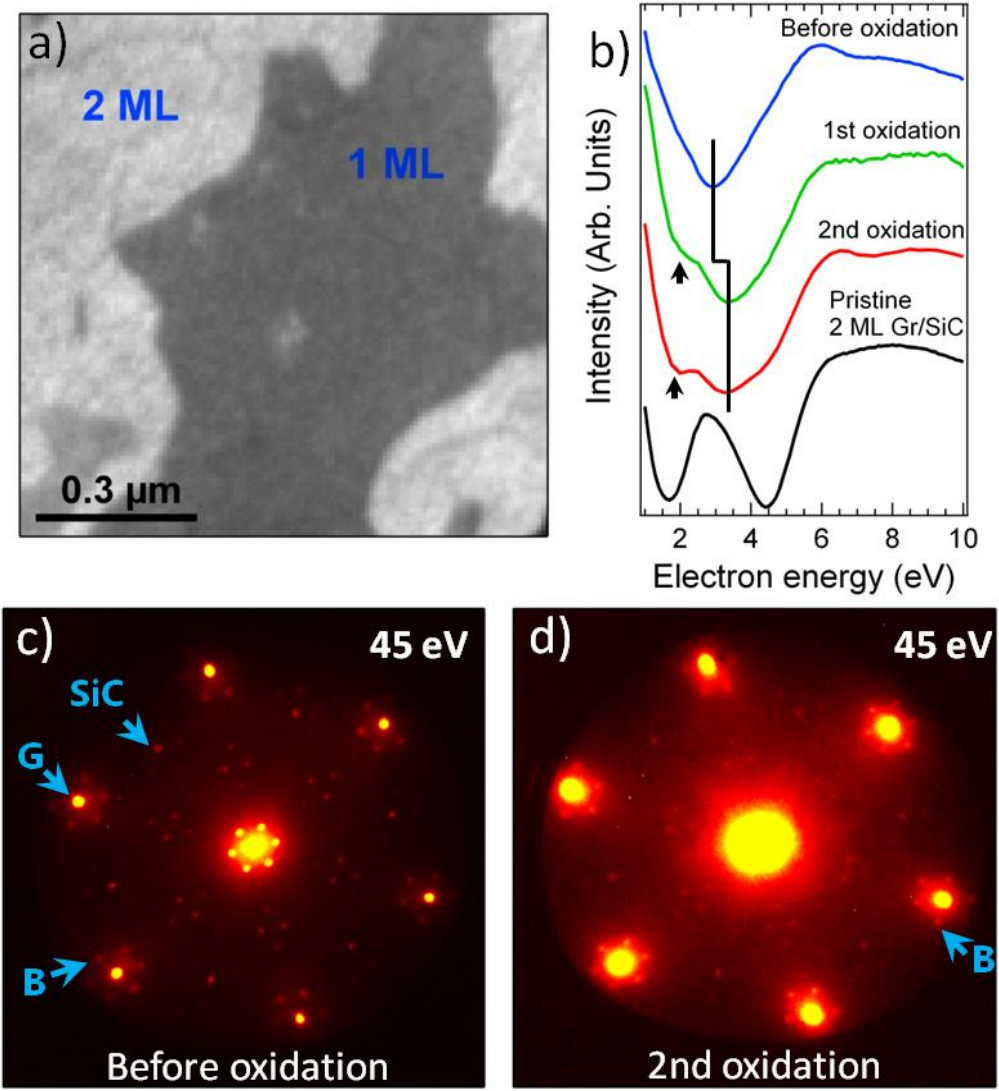

**FIGURE 2**

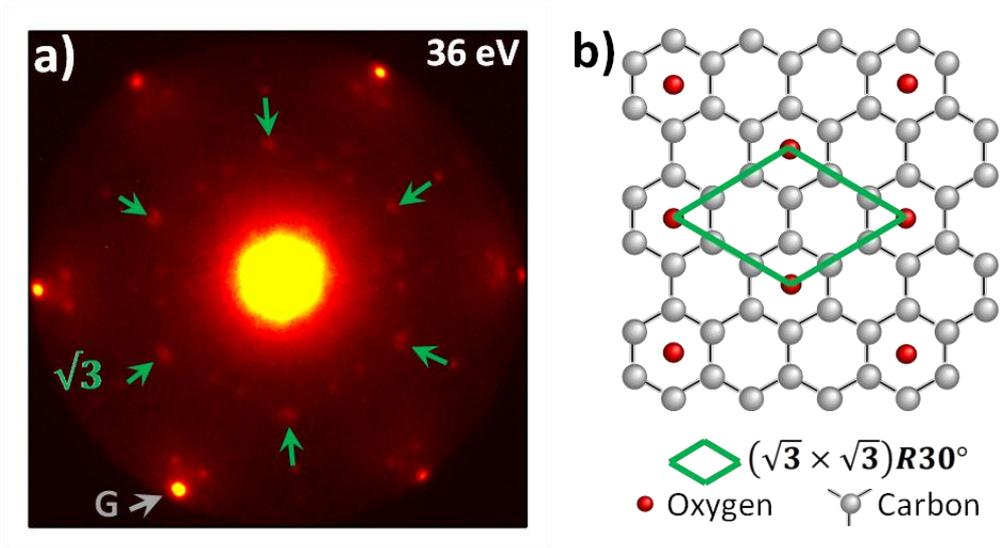

**FIGURE 3**

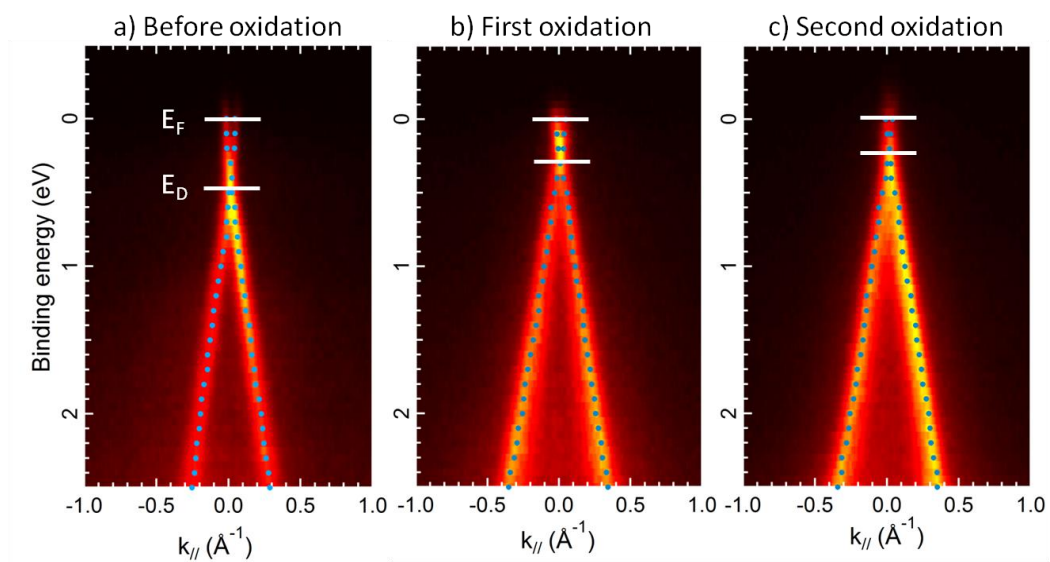

**FIGURE 4**

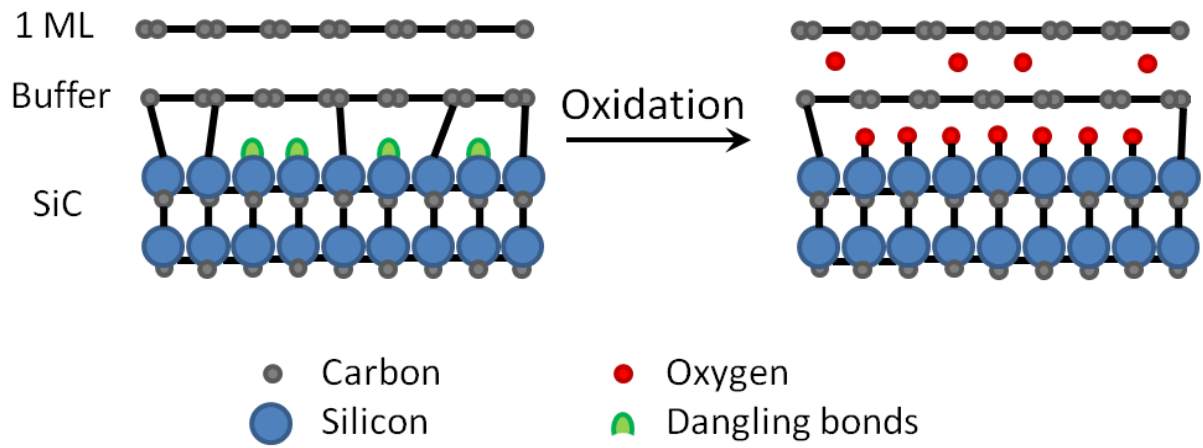

- Carbon
- Silicon
- Oxygen
- Dangling bonds